\documentclass[floatfix,showpacs,prl,twocolumn]{revtex4}
\usepackage{amsbsy,amssymb,amsmath,bm}
\usepackage{graphicx,color}
\usepackage{amsfonts}
\usepackage{amssymb}
\usepackage{amsmath}

\newcommand{\U}[1]{\sigma}

\begin{document}

\title{Dynamics of electric transport in interacting Weyl semimetals.}
\author{B. Rosenstein$^{1,2}$, M. Lewkowicz$^{2}$}
\affiliation{$^{1}$\textit{Electrophysics Department, National Chiao Tung University,
Hsinchu 30050,} \textit{Taiwan, R. O. C}.}
\affiliation{$^{2}$\textit{Physics Department, Ariel University, Ariel 40700, Israel}}
\email{vortexbar@yahoo.com}
\date{\today }

\begin{abstract}
The response to an electric field (DC and AC) of electronic systems in which
the Fermi "surface" consists of a number of 3D Weyl points (such as some
pyrochlore iridates) exhibits a peculiar combination of characteristics
usually associated with insulating and conducting behaviour. Generically a
neutral plasma in clean materials can be described by a tight binding model
with a strong spin-orbit interaction. A system of that type has a vanishing
DC conductivity; however the current response to the DC field is very slow:
the current decays with time in a powerwise manner, different from an
insulator. The AC conductivity, in addition to a finite real part $\sigma
^{\prime }\left( \Omega \right) $ which is linear in frequency, exhibits an
imaginary part $\sigma ^{\prime \prime }\left( \Omega \right) $ \ that
increases logarithmically as function of the UV cutoff (atomic scale). This
leads to substantial dielectric response like a large dielectric constant at
low frequencies. This is in contrast to a 2D Weyl semimetal like graphene at
neutrality point where the AC conductivity is purely pseudo-dissipative. The
Coulomb interaction between electrons is long range and sufficiently strong
to make a significant impact on transport. The interaction contribution to
the AC conductivity is calculated within the tight binding model. The result
for the real part expressed via the renormalized (at frequency $\overline{%
\Omega }$) Fermi velocity $v$ is: $\Delta \sigma ^{\prime }\left( \Omega
\right) =e^{4}\Omega /\left( 9\pi ^{2}\hbar v\right) \left[ 2\log \left(
\Omega /\overline{\Omega }\right) -5\right] $.
\end{abstract}

\maketitle

\section{Introduction and summary\textit{\ }}

A long time ago a rather unorthodox physics of crystals possessing
three-dimensional (3D) pseudo - relativistic quasiparticles \cite{Wolff},
exhibiting an electronic dispersion relation $\varepsilon _{\mathbf{k}%
}=v\left\vert \mathbf{k}\right\vert $, where the velocity $v$ is of the
order of the Fermi velocity in regular condensed matter systems, was invoked
to describe properties of $Bi$. The ultra-relativistic linear dispersion
relation describes two conical bands (of opposite orientation) sharing the
same cone tip. Recently several proposals \cite%
{Ogata,Vishwanath,Kane,Kane,Bernevig} revived an interest in materials with
such excitations nowadays called Weyl semi - metals. The Fermi "surface" of
such materials, typically with dominant spin-orbit interactions, consists
just of a finite number of disconnected points (called Weyl or Dirac points,
defined below) rather than forming a continuous Fermi surface like electrons
in usual metals. The revived interest emerged of course after years of
intense experimental and theoretical study of graphene, a 2D Weyl
quasiparticle material \cite{Katsnelson}. Suspended graphene is just such a
"semimetal" system and exhibits a number of remarkable properties. For
example, despite having zero density of states at the Fermi level and
ideally no impurities, it still has a nonzero DC conductivity\cite{Andrei}.

While this band "touching" at a singular point was noticed in band structure
calculations even before the seminal work of Wallace on graphite\cite%
{Wallace}, the first explicit use of the Dirac model in 3D was in the
context of the two-band approximation model of bismuth\cite{Wolff}. Due to
the strong spin-orbit interaction, the linear in $\mathbf{k}$ terms in a low
energy effective theory near the crystallographic point $L$ of the FCC
Brillouin zone, dominate over quadratic terms (that dominate near the $%
\Gamma $ point, leading to a common "effective mass" description). The
electronic excitations are described by an analog of the Weyl equation of
particle physics, which describes eight two-component chiral spinors $\psi $
(two for each of the four $L$ points)

\begin{equation}
\partial _{t}\psi _{\pm }=\pm v\mathbf{\sigma }\cdot \mathbf{\nabla }\psi
_{\pm }.  \label{Weyl}
\end{equation}%
The sign describes the chirality of the mode. Metallic bismuth is only
approximately described by the ultrarelativistic "massless" dispersion
relation since the quasiparticles of the opposite chirality are coupled and
form four-component massive Dirac bispinors. In $Bi$ therefore electrons are
not Weyl, but $10meV$ massive Dirac electrons where, in addition, the Fermi
level is located away from the Dirac point.

A number of related suggestions for suitable realizations of Weyl semimetals
were recently put forward. Kariyado and Ogata\cite{Ogata} calculated the
band structure of cubic inverse perovskites like $Ca_{3}PbO$ with
significant spin-orbit coupling. They observed the appearance of six Weyl
points with a very small relativistic electron mass down to $4meV$ on the
line connecting the $\Gamma $ - and the $X$ - points. In iridium-based
pyrochlores such as $Y_{2}Ir_{2}O_{7}$, there are $N_{W}=24$ Weyl points
located near the four $L$ points of the FCC lattice. As noted in\cite%
{Vishwanath} these materials "in particular provide a unique opportunity to
study the interplay of Coulomb interactions, spin-orbit coupling, and the
band topology of solids". Also, strong spin-orbit interactions can lead to a
novel phase of matter, the topological insulator\cite{Balents} and various
possibilities to create Weyl fermions combined into coincident opposite
chirality points or separated in the Brillouin zone in $BiO_{2}/SiO_{2}$\cite%
{Kane}, $Na_{3}Bi$,\ $Hg_{1}Cr_{2}Se_{4}$\cite{Bernevig}. These proposals
generated a great deal of experimental efforts\cite{exp3D}. The system with
3D Weyl points was proposed to appear in optical lattices\cite{optlat3D}
following the discovery of "artificial graphene"\cite{optlat2D}.

Since the density of carriers in 3D Weyl semimetals at zero temperature is
zero (as in suspended graphene in 2D), the Coulomb interactions are
unscreened and therefore are expected to be important to the understanding
of the electrical and optical response of these materials. Unsophisticatedly
the dimensionless coupling,%
\begin{equation}
\alpha \equiv \frac{e^{2}}{\epsilon \hbar v},  \label{alfa}
\end{equation}%
is of order $1,$ provided the dielectric constant $\epsilon $ is not large,
since the analog of the light velocity, $v$, is of the order typically of
the Fermi velocity. Note that the same Coulomb potential $1/r$ created by an
electron influences many more electrons in 3D compared to 2D, so naively, in
3D its importance is expected to increase. While electric transport in
noninteracting 3D Weyl fermions was studied (\cite{Ogata12,Vishwanath12} and
references therein to earlier papers in the context of particle physics),
the contributions of potentially very important Coulomb interactions
(Coulomb scattering corrections to transport as an example) have not been
studied theoretically, except basic renormalization effects \cite%
{Vishwanath12}. This is in contrast to the situation in graphene.

The effect of the Coulomb interactions in undoped graphene turned out to be
highly nontrivial, even within perturbation theory, and have evoked a
scientific controversy\cite{Herbut1,Mishchenko,Herbut2,MacDonald} due to the
problem of the "ultraviolet regularization", and was just recently settled%
\cite{Rosenstein13} by noting that the ambiguities are associated with the
treatment of the separation of scales related to the chiral anomaly\cite{Kao}%
. Some aspects of the Weyl semimetal physics are \textit{not} dominated by
the Weyl points of the Brillouin zone at which the spectrum is gapless. For
example, the AC conductivity of undoped graphene (the weak logarithmic
renormalization of the electron velocity \cite{Vozmediano} does not
influence the result), is given in terms of its value in the noninteracting
theory, $\sigma _{0}=e^{2}/4\hbar $, by

\begin{equation}
\frac{\sigma \left( \Omega \right) }{\sigma _{0}}=1+C\alpha +O\left( \alpha
^{2}\right) \text{.}  \label{sigma}
\end{equation}%
and is strictly dissipative (real). This expression is independent of
frequency (provided corrections of order $\hbar \Omega /\gamma $, $\gamma
=2.7eV$ being the hopping energy, are neglected). The value of the only
numerical constant $C$ appearing here has been a matter of intense
controversy. The first detailed calculation\cite{Herbut1} utilizing a sharp
momentum cutoff regularization of the Dirac model provided a value $%
C^{\left( 1\right) }=\frac{25}{12}-\frac{\pi }{2}\approx 0.51$ of order $1$.
The use of the sharp momentum cutoff was criticized by Mishchenko\cite%
{Mishchenko}, who obtained an exceptionally small value of $C^{\left(
2\right) }=\frac{19}{12}-\frac{\pi }{2}\approx 0.01$ making a "soft"
momentum cutoff regularization. He supported this choice by the consistency
of the Kubo and the kinetic equation calculations of conductivity with that
of the polarization function. The consistency required a modification of the
long-range interaction so that it becomes UV cutoff dependent. This
apparently closed the issue. Albeit such a small numerical value would have
profound physical consequences even beyond the transport and dielectric
properties. Nevertheless the interaction strength $C$ was recalculated once
again by Vafek, Juricic and Herbut\cite{Herbut2}, who argued that the
modification of the interaction requires simultaneously a Pauli-Villars
regularization of massless fermions. They applied yet another
regularization, making the space dimensionality fractional, $D=2-\varepsilon
,$ that modified both the current operator and the interaction in such a way
that they satisfy the Ward identities and obtained $C^{\left( 3\right) }=%
\frac{11}{6}-\frac{\pi }{2}\approx 0.26$. The dimensional regularization is
questionable on physical grounds and in a comprehensive subsequent work\cite%
{MacDonald} the authors reaffirmed the small value $C^{\left( 2\right) };$
it seems that this value is the commonly accepted one. The tight binding
calculation\cite{Rosenstein13} however demonstrated that $C^{\left( 3\right)
}$ is the correct one. To reveal the origin of the ambiguity exhibited by
the various values of $C$, the authors made use of a dynamical approach
developed earlier\cite{Lewkowicz} (to address the problem of separating the
interband contributions from the intraband effects due to contacts\cite%
{Ziegler,Lewkowicz11}) directly in the DC case by "switching on" a uniform
electric field in the tight binding model with Coulomb interactions, and
then considering the large-time limit. This approach (known in field theory
as the "infinite hotel story") is the best way to reveal physical effects of
anomalies\cite{Smit}. One can directly separate the contributions from the
neighborhood of Dirac points and the "anomalous" contributions from the rest
of the Brillouin zone, so that one can decide what regularization of the
effective Weyl theory is the correct one. In this sense this is advantageous
over the standard diagrammatic Kubo formula calculation in continuum\cite%
{Mishchenko,Herbut1,Herbut2,MacDonald} that might encounter the so-called
Schwinger terms (found in Quantum Electrodynamics, that is similar to the 3D
Weyl fermions).

The purpose of the present paper is to study the effect of Coulomb
interactions in 3D Weyl fermion systems with emphasis on dynamical aspects
of electric transport and compare/contrast it with the corresponding results
in 2D Weyl fermions. To achieve this goal we define in section II a tight
binding description of Weyl fermions on a hypercubic lattice of any
dimensionality similar to a variant of the Hamiltonian lattice model in
field theory\cite{Smit} already used in its Lagrangian version to simulate
graphene\cite{Drut}. In this model the electron's spin is strongly coupled
to momentum and therefore the model is very reminiscent, in this respect, of
the Wolff model of bismuth or lattice realizations of topological
insulators. The lattice "regularization" is necessary to address the
ultraviolet divergencies at the intermediated stages of calculations, a
problem mentioned above. The universality of this description of Weyl
fermions (of various origins) is supported by the fact that such a 2D model
gives the same result for the interaction corrections as the tight binding
model on the honeycomb lattice with zero spin-orbit coupling.

In Section III the correction to the self energy of a quasiparticle with
momentum $p$ is considered. It is shown that the Fermi velocity
renormalization in 3D (already noted in\cite{Vishwanath12}) is logarithmic
in the UV cutoff $\Lambda \sim \pi /a$ ($a$ being the lattice spacing ) very
much like in 2D\cite{Vozmediano}:%
\begin{equation}
v_{r}\left( p\right) \equiv \frac{\varepsilon _{\mathbf{p}}+\Delta
\varepsilon _{\mathbf{p}}}{p}=v\left[ 1+\frac{2\alpha }{3\pi }\log \left( 
\frac{\hbar \Lambda }{p}\right) \right] \text{.}  \label{RenV}
\end{equation}%
Section IV is devoted to a general derivation and application of the
dynamical approach to the electric response to the DC and AC electric field.
In particular, we obtain the slow current decay in a DC field $E$ of the
neutral, noninteracting 3D Weyl plasma and show that the relaxation is
powerwise, see Fig.1. The long time asymptotics is oscillating and depends
on the microscopic details via cutoff $a$:%
\begin{equation}
\frac{j_{0}\left( t\right) }{E}=\frac{N_{W}e^{2}}{12\pi ^{2}\hbar v}\frac{1}{%
t}\left[ 1+\cos \left( \frac{2vt}{a}\right) \right] \text{.}
\label{cur_decay}
\end{equation}%
Here $N_{W}$ is the number of Weyl fermions. This explains how the
pseudo-Ohmic DC conductivity vanishes in 3D. The relaxation dynamics
therefore is qualitatively different from 2D, where it is insensitive to the
cutoff \cite{Lewkowicz11}. For the AC electric field similar slow
convergence to the AC conductivity occurs, see Fig. 2. Even in the free Weyl
semimetal one gets, in addition to a finite real (pseudo-dissipative) part\
linear in frequency\cite{Vishwanath12}, 
\begin{equation}
\sigma _{0}^{\prime }\left( \Omega \right) =\frac{N_{W}e^{2}}{24\pi \hbar v}%
\Omega \text{,}  \label{sig0}
\end{equation}%
an imaginary part, logarithmically divergent as function of the UV cutoff:%
\begin{equation}
\sigma _{0}^{\prime \prime }\left( \Omega \right) =-\frac{2}{\pi }\sigma
_{0}^{\prime }\left( \Omega \right) \log \frac{\Lambda v}{\Omega }\text{.}
\label{sig0Im}
\end{equation}%
This is again different from graphene at zero doping and leads to important
dielectric properties. Relaxation to the asymptotic behavior is faster at
higher frequencies.

The leading interaction correction to the real and imaginary conductivities
per Weyl point are subject of Section V, see Fig.3 for comparison with a
metal and a semiconductor. The result for the real part expressed via the
renormalized Fermi velocity\ defined still linear in frequency:%
\begin{eqnarray}
\sigma ^{\prime }\left( \Omega \right) &=&\sigma _{0}^{\prime }\left( \Omega
\right) \left[ 1+\,\alpha \left( \frac{2}{3\pi }\log \frac{\Omega }{%
\overline{\Omega }}+C\right) +O\left( \alpha ^{2}\right) \right]
\label{results} \\
C &=&-\frac{5}{3\pi }\approx -0.53\text{.}  \notag
\end{eqnarray}%
The normalization frequency is $\overline{\Omega }=v\overline{p}/\hbar $.
The imaginary part gets further logarithmically dependent on cutoff
corrections,%
\begin{equation}
\sigma ^{\prime \prime }\left( \Omega \right) =\sigma _{0}^{\prime \prime
}\left( \Omega \right) \left[ 1+\frac{\alpha }{3\pi }\log \frac{\Lambda v}{%
\Omega }+O\left( \alpha ^{2}\right) \right] \text{,}  \label{Imsig}
\end{equation}%
despite an apparent "renormalizability" of the model to the two loop order
at least. This is due to the fact that the complex conductivity is not
simply related to basic Green's functions like in the relativistic theory.
The physical significance of the results including formulas including that
for the complex dielectric constant, see Fig.4, are summarized in the
concluding Section VI. We speculate about obvious improvements like the
random phase approximation, renormalization group and a possibility of
stronger coupling effects like the exciton condensation.

\section{Tight binding model with dominant spin-orbit interactions}

\subsection{Noninteracting Hamiltonian and linear response}

The noninteracting tight binding model is defined on the hypercubic lattice $%
\mathbf{n}=\sum_{i=1}^{3}n_{i}\mathbf{a}_{i}$ with lattice vectors $\mathbf{a%
}_{i}$ of length $a$ by the Hamiltonian:

\begin{equation}
K_{mc}=\frac{i}{2}\sum\limits_{\mathbf{n},i}\Gamma _{\mathbf{n},i}c_{\mathbf{%
n}}^{\dagger }\sigma _{i}c_{\mathbf{n}+\mathbf{a}_{i}}+hc\text{.}
\label{Kdef}
\end{equation}%
Here\ $\sigma _{i}$ are Pauli matrices, operators $c_{\mathbf{n}}^{\alpha
\dagger }$ create a two component spinor $\alpha =1,2$ and $\Gamma _{\mathbf{%
n},i}$ is the hopping integral that in the presence of an external
electromagnetic field, described by vector potential $A_{i},$ is 
\begin{equation}
\Gamma _{\mathbf{n},i}=\gamma \exp \left[ i\frac{ea}{c\hbar }%
\int_{s=0}^{1}A_{i}\left( \mathbf{n}+s\mathbf{a}_{i},t\right) \right] ,
\label{Wilson}
\end{equation}%
where the hopping energy $\gamma $ is of order of the band width. It is
important to derive the current density directly in tight binding model: 
\begin{eqnarray}
J_{i}\left( r,t\right) &\equiv &-c\frac{\partial K_{mc}}{\partial
A_{i}\left( \mathbf{r},t\right) }  \label{Jdef} \\
&=&\frac{ea}{2\hbar }\sum\limits_{n}\int_{s=0}^{1}\delta ^{D}\left( \mathbf{r%
}-\mathbf{n}-s\mathbf{a}_{i}\right) \Gamma _{\mathbf{n},i}c_{\mathbf{n}%
}^{\dagger }\sigma _{i}c_{\mathbf{n}+\mathbf{a}_{i}}+hc.  \notag
\end{eqnarray}%
It defines the UV regularization of the current operator that obeys the Ward
identities. In linear response the current density operator is expanded up
to the first order in $\mathbf{A}$ as $\mathbf{J}=\mathbf{J}_{p}+\mathbf{J}%
_{d}$:

\begin{eqnarray}
J_{i}^{p}\left( \mathbf{r}\right) &=&\frac{ev}{2}\sum\limits_{n}\Upsilon
_{n}+hc,  \label{JpJd_def} \\
J_{i}^{d}\left( \mathbf{r},t\right) &=&i\frac{e^{2}va}{2c\hbar }A_{i}\left(
r,t\right) \sum\limits_{n}\Upsilon _{n}+hc,  \notag \\
\Upsilon _{n} &=&\int_{s=0}^{1}\delta ^{D}\left( \mathbf{r}-\mathbf{n}-s%
\mathbf{a}_{i}\right) c_{\mathbf{n}}^{\dagger }\sigma _{i}c_{\mathbf{n}+%
\mathbf{a}_{i}}
\end{eqnarray}%
where $v=\gamma a/\hbar $. Space averages over volume $\mathcal{V}$ for a
homogeneous vector potential $\mathbf{A}\left( t\right) $ simplify:%
\begin{eqnarray}
j_{i}^{p} &=&\frac{1}{\mathcal{V}}\int_{\mathbf{r}}J_{i}^{p}\left( \mathbf{r}%
\right) =\frac{ev}{2\mathcal{V}}\sum\limits_{\mathbf{n}}c_{\mathbf{n}%
}^{\dagger }\sigma _{i}c_{\mathbf{n}+\mathbf{a}_{i}}+hc,  \label{gpgd} \\
j_{i}^{d} &=&\frac{1}{\mathcal{V}}\int_{\mathbf{r}}J_{i}^{d}\left( \mathbf{r}%
,t\right) =i\frac{e^{2}av}{2c\hbar \mathcal{V}}A_{i}\left( t\right)
\sum\limits_{\mathbf{n}}c_{\mathbf{n}}^{\dagger }\sigma _{i}c_{\mathbf{n}+%
\mathbf{a}_{i}}+hc  \notag
\end{eqnarray}%
Expansion of the minimally coupled Hamiltonian in electric field is

\begin{equation}
K_{mc}\approx K+H_{ext}\text{ ; \ \ \ }H_{ext}=-\frac{1}{c}\int_{\mathbf{r}}%
\mathbf{J}^{p}\cdot \mathbf{A}\left( t\right) .  \label{Ahomogen}
\end{equation}

Defining Fourier components by $c_{\mathbf{n}}^{\alpha }=\mathcal{N}%
^{-1/2}\sum\limits_{\mathbf{k}}e^{-i\mathbf{k}\cdot \mathbf{n}}c_{\mathbf{k}%
}^{\alpha };$ where the number of unit cells is $\mathcal{N}=\mathcal{V}%
/a^{D},$ one has:%
\begin{equation}
K=\gamma \sum\limits_{\mathbf{k},i}\sin \left( k_{i}a\right) c_{\mathbf{k}%
}^{+}\sigma _{i}c_{\mathbf{k}}.  \label{FourierK}
\end{equation}%
Diagonalization of $K$ is achieved adopting a reinterpretation of the
absence of an electron in the valence band as a hole and using units $\hbar
=a=v=1$ by

\begin{equation}
c_{\mathbf{k}}^{\alpha }=v_{\mathbf{k}}^{\alpha }a_{\mathbf{k}}+u_{\mathbf{k}%
}^{\alpha }b_{-\mathbf{k}}^{+}\rightarrow c_{\mathbf{k}}^{\alpha +}=v_{%
\mathbf{k}}^{\alpha \ast }a_{\mathbf{k}}^{+}+u_{\mathbf{k}}^{\alpha \ast
}b_{-\mathbf{k}},  \label{Bogtrans}
\end{equation}%
with spinors $v_{\mathbf{k}}$ and $u_{\mathbf{k}}$ given in Appendix A. Up
to an additive constant it becomes

\begin{equation}
K=\sum\nolimits_{\mathbf{k}}\varepsilon _{\mathbf{k}}\left( a_{\mathbf{k}%
}^{+}a_{\mathbf{k}}+b_{\mathbf{k}}^{+}b_{\mathbf{k}}\right) ,
\label{Ksecond}
\end{equation}%
where 
\begin{equation}
\varepsilon _{\mathbf{k}}=\sqrt{\widehat{k_{1}}^{2}+\widehat{k_{2}}^{2}+%
\widehat{k_{3}}^{2}}\text{,}  \label{epsilon}
\end{equation}%
and the notation $\widehat{k}\equiv \sin k$ was introduced. In 3D one
observes 8 Weyl points at which $\varepsilon _{\mathbf{k}}=0$ inside the
Brillouin zone (BZ). Four are right handed: one in the center (the $\Gamma $
point) and three on the faces ($X$), while four are left handed: three on
the edges ($M$) and one in the corner ($R$). The chirality is determined by
the expansion of the Hamiltonian Eq.(\ref{FourierK}) around a Weyl point $%
\mathbf{W}$, $sgn(\varepsilon _{ijk}Q_{i}^{1}Q_{j}^{2}Q_{k}^{3})$, where $%
\mathbf{Q}^{i}=\frac{\partial }{\partial \mathbf{k}}\widehat{k_{i}}|_{%
\mathbf{W}}$. Similarly, in 2D one has two right handed Weyls at $\Gamma $
and $R$ and two left handed at $M$.

The paramagnetic and diamagnetic parts of the current due to an electric
field (along, let's say, the $z$ direction), $\mathbf{A}\left( t\right)
=\left( 0,0,A\left( t\right) \right) ;$ $E\left( t\right) =-c\frac{d}{dt}%
A\left( t\right) $, are:%
\begin{widetext}%
\begin{eqnarray}
j^{p} &=&\frac{e}{\mathcal{V}}\sum\limits_{\mathbf{k}}\left[ \iota _{\mathbf{%
k}}\left( a_{\mathbf{k}}^{+}a_{\mathbf{k}}+b_{-\mathbf{k}}^{+}b_{-\mathbf{k}%
}\right) +i\chi _{\mathbf{k}}b_{-\mathbf{k}}a_{\mathbf{k}}\right] +hc;
\label{gpgddef} \\
j^{d} &=&\frac{e^{2}}{\mathcal{V}}A\left( t\right) \sum\limits_{\mathbf{k}}%
\frac{\widehat{k_{z}}}{\varepsilon _{\mathbf{k}}}\left[ \frac{\widehat{k_{z}}%
}{2}\left( a_{\mathbf{k}}^{+}a_{\mathbf{k}}+b_{-\mathbf{k}}^{+}b_{-\mathbf{k}%
}\right) -\left( \widehat{k_{x}}+i\widehat{k_{y}}\right) b_{-\mathbf{k}}a_{%
\mathbf{k}}\right] +hc,  \notag
\end{eqnarray}%
\end{widetext}%
where the functions $\chi _{\mathbf{k}}$ and $\iota _{\mathbf{k}}$ are
defined in Appendix A.

\subsection{Coulomb interaction}

Using the expression for the particle densities in momentum space (Einstein
summation implied for spins only)

\begin{equation}
N_{\mathbf{n}}=c_{\mathbf{n}}^{\alpha \dagger }c_{\mathbf{n}}^{\alpha }=%
\mathcal{N}^{-1}\sum\limits_{\mathbf{kl}}e^{i\left( \mathbf{l}-k\right)
\cdot \mathbf{n}}c_{\mathbf{l}}^{\alpha +}c_{\mathbf{k}}^{\alpha }\text{,}
\label{density}
\end{equation}%
the Coulomb interaction takes the form $\ $

\begin{equation}
V=\frac{1}{2}\sum\limits_{\mathbf{nm}}\frac{e^{2}}{\left\vert \mathbf{n}-%
\mathbf{m}\right\vert }N_{\mathbf{n}}N_{\mathbf{m}}=\sum\limits_{\mathbf{klk}%
^{\prime }\mathbf{l}^{\prime }}v_{\mathbf{klk}^{\prime }\mathbf{l}^{\prime
}}c_{\mathbf{l}}^{\alpha +}c_{\mathbf{k}}^{\alpha }c_{\mathbf{l}^{\prime
}}^{\alpha ^{\prime }+}c_{\mathbf{k}^{\prime }}^{\alpha ^{\prime }}\text{;}
\label{Coulomb_c}
\end{equation}%
\begin{equation}
v_{\mathbf{klk}^{\prime }\mathbf{l}^{\prime }}=\frac{1}{2\mathcal{V}}\sum_{%
\mathbf{p}}v_{\mathbf{p}}\delta _{\mathbf{l-k-p}}\delta _{\mathbf{l}^{\prime
}\mathbf{-k}^{\prime }\mathbf{+p}},  \label{vdef}
\end{equation}%
with Fourier transform of the interaction being $v_{\mathbf{p}}=\frac{4\pi
e^{2}}{a^{2}\mathbf{p}^{2}}$. It is important to note that the electric
charge of the effective tight binding model should include the contributions
to the screening due to the polarization constant $\epsilon $ caused by
degrees of freedom not included in the model, so the "bare" charge includes
this effect $e^{2}=e_{el}^{2}/\epsilon $. We require charge neutrality that
is achieved by leaving out all the contributions including $v_{\mathbf{p=0}}$%
. This prescription is always implied in what follows. It is convenient to
normal order $V$ \ with respect to\ operators $a$ and $b$ diagonalizing the
"kinetic" term $K$ via Eq.(\ref{Bogtrans}):%
\begin{equation}
V=V^{40}+V^{31}+V^{22}+V^{13}+V^{04}+V^{11}\text{,}  \label{parts}
\end{equation}%
where the part $V^{ij}$ contains $i$ creation operators $a^{+}$ or $b^{+}$
and $j$ annihilation operators $a$ or $b$, all specified in Appendix A. In
principle the quadratic pair creation $V^{20}$ and the pair annihilation
terms $V^{02}$ a could have appeared. The fact that they have not, explained
in the Appendix A, greatly simplifies the calculation and makes it
competitive (at least to the two-loop order) with the diagrammatic approach.
The "time independent" approach is however much more transparent, when one
realizes that excitations can be created in fours rather than in pairs in
this particular model.

\section{Quasiparticles and renormalization of the Fermi velocity}

The energy of an electron above the Fermi level with quasimomentum $\mathbf{p%
}$, $\left\vert \mathbf{p}\right\rangle =a_{\mathbf{p}}^{+}\left\vert
0\right\rangle $, in the noninteracting model described by the Hamiltonian
Eq.(\ref{K}) is:

\begin{equation}
\left\langle \mathbf{p}\left\vert K\right\vert \mathbf{p}\right\rangle
=\left\langle 0\left\vert a_{\mathbf{p}}\sum\nolimits_{\mathbf{l}%
}\varepsilon _{\mathbf{l}}\left( a_{\mathbf{l}}^{+}a_{\mathbf{l}}+b_{\mathbf{%
l}}^{+}b_{\mathbf{l}}\right) a_{\mathbf{p}}^{+}\right\vert 0\right\rangle
=\varepsilon _{\mathbf{p}}\text{.}  \label{qpenergy0}
\end{equation}%
The interaction correction is%
\begin{equation}
\Delta \varepsilon _{\mathbf{p}}=\left\langle \mathbf{p}\left\vert
V\right\vert \mathbf{p}\right\rangle =\left\langle 0\left\vert a_{\mathbf{p}%
}Va_{\mathbf{p}}^{+}\right\vert 0\right\rangle \text{.}  \label{qpenergy1}
\end{equation}%
Obviously only contributions with equal numbers of creation and annihilation
operators, $V^{22}$ and $V^{11}$ can contribute. The first term contains
(see Appendix A) $a^{+}b^{+}ab$, $b^{+}b^{+}bb$ and $a^{+}a^{+}aa$ terms, of
which only the last one could contribute to the expectation value Eq.(\ref%
{qpenergy1}), yet it vanishes:%
\begin{eqnarray}
&&\Delta \varepsilon _{\mathbf{p}}^{22}  \label{deleps22} \\
&=&-\sum\limits_{\mathbf{klk}^{\prime }\mathbf{l}^{\prime }}v_{\mathbf{klk}%
^{\prime }\mathbf{l}^{\prime }}\left( v_{\mathbf{l}^{\prime }}^{\ast }\cdot
v_{\mathbf{k}^{\prime }}\right) \left( v_{\mathbf{l}}^{\ast }\cdot v_{%
\mathbf{k}}\right) \left\langle 0\left\vert a_{\mathbf{p}}a_{\mathbf{l}%
}^{+}a_{\mathbf{l}^{\prime }}^{+}a_{\mathbf{k}}a_{\mathbf{k}^{\prime }}a_{%
\mathbf{p}}^{+}\right\vert 0\right\rangle  \notag \\
&=&-\sum\limits_{\mathbf{kll}^{\prime }}v_{\mathbf{klpl}^{\prime }}\left( v_{%
\mathbf{l}^{\prime }}^{\ast }\cdot v_{\mathbf{p}}\right) \left( v_{\mathbf{l}%
}^{\ast }\cdot v_{\mathbf{k}}\right) \left\langle 0\left\vert a_{\mathbf{p}%
}a_{\mathbf{l}}^{+}a_{\mathbf{l}^{\prime }}^{+}a_{\mathbf{k}}\right\vert
0\right\rangle =0\text{.}  \notag
\end{eqnarray}

We are left with the simple quadratic part $V^{11}$:%
\begin{eqnarray}
\Delta \varepsilon _{\mathbf{p}}^{11} &=&\frac{1}{2}\sum\limits_{\mathbf{ql}%
}v_{\mathbf{l-q}}g_{\mathbf{ql}}\left\langle 0\left\vert a_{\mathbf{p}%
}\left( a_{\mathbf{l}}^{+}a_{\mathbf{l}}+b_{-\mathbf{l}}^{+}b_{-\mathbf{l}%
}\right) a_{\mathbf{p}}^{+}\right\vert 0\right\rangle  \notag \\
&=&\frac{1}{2}\sum\limits_{\mathbf{q}}v_{\mathbf{p-q}}g_{\mathbf{qp}}\text{,}
\label{deleps11new}
\end{eqnarray}%
where%
\begin{equation}
g_{\mathbf{pq}}\equiv \left\vert v_{\mathbf{q}}^{\ast }\cdot v_{\mathbf{p}%
}\right\vert ^{2}-\left\vert u_{\mathbf{q}}^{\ast }\cdot v_{\mathbf{p}%
}\right\vert ^{2}\text{.}  \label{Aqp}
\end{equation}%
This results in:%
\begin{equation}
\Delta \varepsilon _{\mathbf{p}}=\frac{e^{2}p}{3\pi }\left[ \log \frac{3\pi
^{2}}{2a^{2}p^{2}}+O\left( p^{2}\right) \right] \text{.}
\label{deleps_exact}
\end{equation}

It is instructive to estimate it using the expansion around any of the eight
Weyl points with the momenta restricted, say, by $q<\Lambda <\pi /2a$, thus
omitting some contributions far from the Weyl points that innocently could
be thought to be small, but, as the graphene example taught us, might become
perfidious if powerwise UV divergencies appear in the intermediate stages of
the calculation. In the present case divergencies are just logarithmic
though and one proceeds by expanding around\ one of the Weyl points,

\begin{equation}
g_{\mathbf{pq}}\approx \cos \theta _{\mathbf{q}}\cos \theta _{\mathbf{p}%
}+\sin \theta _{\mathbf{q}}\sin \theta _{\mathbf{p}}\cos \left( \phi _{%
\mathbf{p}}-\phi _{\mathbf{q}}\right) \text{,}  \label{AqpWeyl}
\end{equation}%
where spherical coordinates, $\mathbf{q=}q\left( \sin \theta _{\mathbf{q}%
}\cos \phi _{\mathbf{q}},\sin \theta _{\mathbf{q}}\sin \phi _{\mathbf{q}%
},\cos \theta _{\mathbf{q}}\right) $, were used. The sum is transformed into
an integral:%
\begin{equation}
\Delta \varepsilon _{\mathbf{p}}^{11}=\frac{e^{2}}{2\left( 2\pi \right) ^{3}}%
\int_{q\phi _{\mathbf{q}}\theta _{\mathbf{q}}}\frac{4\pi \sin \theta _{%
\mathbf{q}}q^{2}g_{\mathbf{pq}}}{p^{2}+q^{2}-2pqg_{\mathbf{pq}}}\text{.}
\label{depsWeyl}
\end{equation}%
The integrals over angles can be performed, see Appendix B, resulting in
(for $\mathbf{p=}p\left( 1,0,0\right) $) 
\begin{equation}
\Delta \varepsilon _{\mathbf{p}}^{11}=\frac{e^{2}}{4\pi }\int_{q=0}^{\Lambda
}\left[ \frac{1+r^{2}}{2}\log \left( \frac{1+r}{1-r}\right) ^{2}-2r\right] 
\text{,}  \label{deps1}
\end{equation}%
where $r=q/p$. Changing variables in the remaining integral establishes the
leading linear dependence on $p$: 
\begin{eqnarray}
\Delta \varepsilon _{\mathbf{p}} &=&\frac{e^{2}p}{4\pi }\int_{r=0}^{\Lambda
/p}\left[ \frac{1+r^{2}}{2}\log \left( \frac{1+r}{1-r}\right) ^{2}-2r\right]
\label{delepsfinal} \\
&=&\frac{e^{2}p}{3\pi }\left[ \log \left( \frac{\Lambda ^{2}}{p^{2}}\right) +%
\frac{5}{3}\right] \text{.}  \notag
\end{eqnarray}%
Comparing with the exact result for the "universal" model, Eq.(\ref%
{deleps_exact}), we can choose the value of the cutoff to be slightly
outside of the continuum model applicability range: $\Lambda =0.53\frac{\pi 
}{a}$ . This means that the continuum model cannot determine correctly the
constant term that is model dependent. To summarize, one indeed observes
just a logarithmic divergence and therefore can renormalize the Fermi
velocity written in physical units in Eq.(\ref{RenV}).

In 2D the corresponding calculation gives a well known "running" of the
graphene velocity towards higher velocities\cite{Vozmediano}. Therefore the
running of the Fermi velocity in Weyl semimetal is very much like in
(undoped) graphene.

\section{Dynamics of electric response: current decay in DC field and
inductive response in AC field of the noninteracting Weyl fermions neutral
plasma.}

In this section we first calculate the time evolution of the current when
the external electric field is switched on in order to obtain the DC and AC
electric response. This is done however within a "time independent"
formalism to carefully trace the states that contribute to the dynamics and
investigate the emergence of the steady state.

\subsection{The electric current evolution within linear response}

We use the dynamical approach\cite{Lewkowicz} to the semimetal response
rather than the more customary diagrammatic method not just to investigate
dynamics under DC or AC field, but also to clarify several fundamental
issues in the following sections concerning the Coulomb interaction
corrections to the AC conductivity. First the nature of the dependence of
the physical quantities on the UV "cutoff" $1/a$ (renormalization) is
elucidated, next the time evolution is exploited to demonstrate the
inductive response of the 3D semimetal even without electron-electron
interaction and contrast it with the purely pseudo-Ohmic response of
graphene.

The paramagnetic contribution in the Heisenberg picture is given by the
vacuum expectation value (VEV):

\begin{eqnarray}
\left\langle \mathbf{J}^{p}\left( \mathbf{r}\right) \right\rangle
&=&\left\langle \psi _{mc}\left( t\right) \left\vert \mathbf{J}^{p}\left( 
\mathbf{r}\right) \right\vert \psi _{mc}\left( t\right) \right\rangle
\label{para} \\
&=&\left\langle \psi \left( 0\right) \left\vert U_{mc}^{-1}\left( t\right) 
\mathbf{J}^{p}\left( \mathbf{r}\right) U_{mc}\left( t\right) \right\vert
\psi \left( 0\right) \right\rangle \text{.}  \notag
\end{eqnarray}%
The ground state $\psi \left( 0\right) $ is that of an interacting electron
system without the external field. Expanding to "linear response" in
coupling to the external electric field (considered homogeneous and oriented
along the $z$ direction), Eq.(\ref{Ahomogen}),%
\begin{equation}
U_{mc}\left( t\right) =U_{K}\left( t\right) \left[ 1-i%
\int_{t_{1}=0}^{t}U_{H}^{-1}\left( t_{1}\right) H_{ext}\left( t_{1}\right)
U_{H}\left( t_{1}\right) \right] \text{,}  \label{Umc}
\end{equation}%
where $U_{H}\left( t\right) =e^{-i\left( K+V\right) t}$, one gets%
\begin{widetext}%

\begin{equation}
\left\langle \mathbf{J}^{p}\left( \mathbf{r,}t\right) \right\rangle
=i\int_{t_{1}=0}^{t}A\left( t_{1}\right) \int_{r^{\prime }}\left\langle \psi
\left( 0\right) \left\vert e^{iHt}J_{z}^{p}\left( \mathbf{r}\right)
e^{-iH\left( t-t_{1}\right) }\mathbf{J}^{p}\left( \mathbf{r}^{\prime
}\right) e^{-iHt_{1}}\right\vert \psi \left( 0\right) \right\rangle +cc\text{%
.}  \label{Jplinresp}
\end{equation}%
\end{widetext}%
The diamagnetic current, Eq.(\ref{JpJd_def}) is already of the first order
in electric field. In the next section the full model including the many
body effects will be considered. In this section we neglect the interaction $%
V$ to solve exactly for the time evolution of the current density.

\subsection{Time dependence of the current generated by a time dependent
electric field}

In the absence of the Coulomb interaction the VEV of the average
paramagnetic current density (in the field direction $z$, see Eq.(\ref%
{JpJd_def})) takes a form:

\begin{eqnarray}
&&\left\langle j_{0}^{p}\left( t\right) \right\rangle  \label{gpnonint} \\
&=&i\mathcal{V}\int_{t_{1}=0}^{t}A\left( t_{1}\right) \left\langle
0\left\vert e^{iKt}j^{p}e^{-iK\left( t-t_{1}\right)
}j^{p}e^{-iKt_{1}}\right\vert 0\right\rangle +cc\text{.}  \notag
\end{eqnarray}%
Using definitions of the tight binding Hamiltonian and the current density
operator, Eqs. (\ref{Ksecond}) and (\ref{gpgddef}), one obtains

\begin{equation}
\left\langle j_{0}^{p}\left( t\right) \right\rangle =\frac{e^{2}}{\mathcal{V}%
}\int_{t_{1}=0}^{t}A\left( t_{1}\right) \sum\limits_{\mathbf{k}}\left\vert
\chi _{\mathbf{k}}\right\vert ^{2}\sin \left[ 2\varepsilon _{\mathbf{k}%
}\left( t-t_{1}\right) \right] \text{.}  \label{gpfree}
\end{equation}%
Similarly the diamagnetic contribution, using Eq.(\ref{gpgddef}), is

\begin{equation}
\left\langle j_{0}^{d}\left( t\right) \right\rangle =\left\langle
0\left\vert j^{d}\right\vert 0\right\rangle =-\frac{e^{2}}{\mathcal{V}}%
A\left( t\right) \sum\limits_{\mathbf{k}}\frac{\widehat{k}_{z}^{2}}{%
\varepsilon _{k}}.  \label{gdfree}
\end{equation}%
To continue one has to specify the time dependence of the applied electric
field. We start in the next subsection with the constant electric field and
then continue to the AC case.

\subsection{Decay of current in the DC electric field}

In the homogeneous DC electric field described by the vector potential $%
A\left( t\right) =-cEt$, one obtains from Eq.(\ref{gpfree}) the average
current. The integral over $t_{1}$ results in:

\begin{equation}
\frac{j_{0}\left( t\right) }{E}=\frac{e^{2}}{\mathcal{V}}\sum\limits_{%
\mathbf{k}}\left\{ t\frac{\widehat{k}_{z}^{2}}{\varepsilon _{\mathbf{k}}}%
-\left\vert \chi _{\mathbf{k}}\right\vert ^{2}\left[ \frac{t}{2\varepsilon _{%
\mathbf{k}}}-\frac{\sin \left( 2\varepsilon _{\mathbf{k}}t\right) }{%
4\varepsilon _{\mathbf{k}}^{2}}\right] \right\} \text{.}  \label{sigmaDC}
\end{equation}%
One expects that the neutral plasma system that does not possess electric
charges on the Fermi level will not have the "acceleration" terms linear in
time $t$ that appear in the above equation. Indeed the sum of the first two
terms vanishes:%
\begin{equation}
\frac{j_{0}^{acc}\left( t\right) }{E}=-t\frac{e^{2}}{\mathcal{V}}%
\sum\limits_{\mathbf{k}}\left[ \varepsilon _{\mathbf{k}}^{-3}\left( 1-%
\widehat{k}_{z}^{2}\right) \left( \widehat{k}_{x}^{2}+\widehat{k}%
_{y}^{2}\right) +\varepsilon _{\mathbf{k}}^{-1}\widehat{k}_{z}^{2}\right] =0.
\label{sigmaacc}
\end{equation}%
This is seen as follows. The integral over the BZ can be represented as an
integral over the full $\frac{d}{dk_{z}}$derivative like in graphene\cite%
{Lewkowicz}. Since the BZ can be taken periodic in the quasimomentum
component $k_{z}$, the integral over the derivative vanishes. The physical
arguments put forward in the framework of graphene\cite{Kao} in order to
comprehend this apply equally well here and are not repeated. The remaining
non-accelerating part,%
\begin{equation}
\frac{j_{0}\left( t\right) }{E}=\frac{e^{2}}{2\mathcal{V}}\sum\limits_{%
\mathbf{k}}\varepsilon _{\mathbf{k}}^{-4}\left( 1-\widehat{k}_{z}^{2}\right)
\left( \widehat{k}_{x}^{2}+\widehat{k}_{y}^{2}\right) \sin \left(
2\varepsilon _{\mathbf{k}}t\right) ,  \label{jdyn}
\end{equation}%
is presented in Fig.1. It decays as given (in physical units) in Eq.(\ref%
{cur_decay}),\ exhibiting the zero conductivity rather than a universal
finite value as in graphene. This asymptotic value of zero is approached
therefore powerwise and oscillating. Note the dependence on the ultraviolet
cutoff, that did not appear in graphene\cite{Lewkowicz11}.

\begin{figure}[tbp]
\begin{center}
\includegraphics[width=6cm]{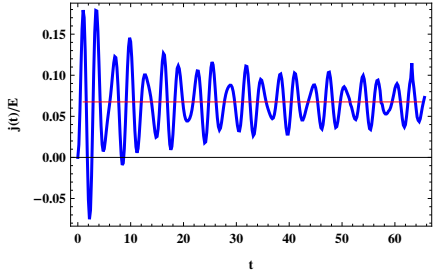}
\end{center}
\caption{Fig.1 The response of the free
Weyl semimetal to a DC electric field, multiplied by $t,$ is shown as a
function of time. The current's asymptotic value of zero is approached
powerwise.}
\end{figure}

\subsection{Dissipative and inductive response to the AC electric field}

The AC electric field, $E\cos \left( \Omega t\right) $, is represented by
the oscillating vector potential, $A\left( t\right) =-\frac{cE}{\Omega }\sin
\left( \Omega t\right) $. The "universal" nearest neighbours tight binding
model is relevant only for frequencies $\Omega $ smaller than the hopping
integral $\gamma /\hbar $. However, as we noticed in the context of graphene%
\cite{Kao}, the use of the expansion of the dispersion relation near the
Weyl points should be done with some care.

Performing the integral over $t_{1}$ in Eq.(\ref{gpfree}), one now obtains
the current:%
\begin{widetext}%
\begin{equation}
\frac{j_{0}\left( t\right) }{E}=\frac{e^{2}}{\mathcal{V}\Omega }\sum\limits_{%
\mathbf{k}}\left\{ \varepsilon _{\mathbf{k}}^{-1}\widehat{k}_{3}^{2}\sin
\left( \Omega t\right) -2\varepsilon _{\mathbf{k}}^{-2}\left( 1-\widehat{k}%
_{z}^{2}\right) \left( \widehat{k}_{x}^{2}+\widehat{k}_{y}^{2}\right) \frac{%
\Omega \sin \left( 2\varepsilon _{\mathbf{k}}t\right) -2\varepsilon _{%
\mathbf{k}}\sin \left( \Omega t\right) }{\Omega ^{2}-4\varepsilon _{\mathbf{k%
}}^{2}}\right\} \text{.}  \label{sigmaAC}
\end{equation}%
\end{widetext}%
As in graphene, it is very difficult to approach frequencies of interest $%
\Omega <<\gamma /\hbar $, when the conductivity is presented in this form.
However it becomes substantially simpler numerically when one subtracts the
vanishing acceleration term Eq.(\ref{sigmaacc}), that we have encountered in
the DC calculation, with $t$ replaced by $1/\Omega $:%
\begin{widetext}%
\begin{equation}
\frac{j_{0}\left( t\right) }{E}=\frac{e^{2}}{\mathcal{V}}\sum\limits_{%
\mathbf{k}}\varepsilon _{\mathbf{k}}^{-3}\left( 1-\widehat{k}_{z}^{2}\right)
\left( \widehat{k}_{x}^{2}+\widehat{k}_{y}^{2}\right) \frac{\Omega \sin
\left( \Omega t\right) -2\varepsilon _{\mathbf{k}}\sin \left( 2\varepsilon _{%
\mathbf{k}}t\right) }{\Omega ^{2}-4\varepsilon _{\mathbf{k}}^{2}}\text{.}
\label{sigma0}
\end{equation}%
\end{widetext}%
This is shown (together with the electric field) in Fig.2 for several values
of frequency. One observes that beyond certain relaxation time the response
becomes periodic exhibiting a phase difference between the current (points)
and the electric field (solid lines). To obtain the steady state value of
the complex conductivity we average over time with a damping factor $\eta $
using%
\begin{equation}
\sigma \left( \Omega \right) =\underset{\eta \rightarrow 0+}{\lim }\frac{%
2\eta }{E}\int_{t=0}^{\infty }e^{i\Omega t-\eta t}j\left( t\right) \text{.}
\label{etatrick}
\end{equation}%
The time integration in Eq.(\ref{sigma0}) with the current density of Eq.(%
\ref{sigma0}) results in 
\begin{equation}
\sigma _{0}\left( \Omega \right) =-\frac{ie^{2}\Omega }{\mathcal{V}}%
\sum\limits_{\mathbf{k}}\varepsilon _{\mathbf{k}}^{-3}\left( 1-\widehat{k}%
_{z}^{2}\right) \left( \widehat{k}_{x}^{2}+\widehat{k}_{y}^{2}\right) \frac{1%
}{4\varepsilon _{\mathbf{k}}^{2}-\Omega _{+}^{2}}\text{,}  \label{sigma_eta}
\end{equation}%
where $\Omega _{+}\equiv \Omega +i\eta $. The exact integral is presented
(in physical units) in Eq.(\ref{sig0}) and Eq.(\ref{sig0Im}) with
nonuniversal cutoff related to the lattice spacing via $\Lambda =0.73\frac{%
\pi }{a}$.

\begin{figure}[tbp]
\begin{center}
\includegraphics[width=6cm]{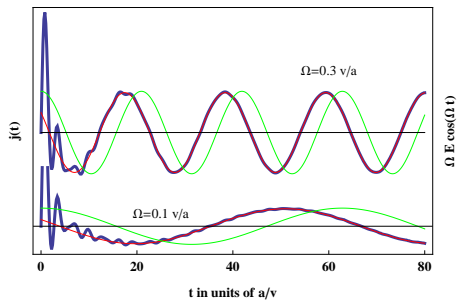}
\end{center}
\caption{Fig.2 The response of the free
Weyl semimetal to an AC electric field (green lines) is shown as a function
of time for two different frequencies (magenta lines). The value of $v/a$ is
typically of order $\sim 3\cdot 10^{15}Hz.$}
\end{figure}

Now we proceed to try to use the continuum model near the Weyl points. After
the subtraction of "anomalous" acceleration terms most of the contributions
for $\Omega <<\gamma /\hbar $ come from the immediate neighbourhoods of the
Weyl points\cite{Lewkowicz}. Due to symmetries it suffices to consider one
of them. The neighbourhood of the origin will be defined by $k<\Lambda <\pi
/2$. Here two differences emanate compared to the transport in graphene. In
graphene the lattice spacing is irrelevant, so that the conductivity can be
calculated within the continuum Weyl model and the AC conductivity is real
for all frequencies $\Omega <<\gamma /\hbar $. In the present case an
inductive part appears and moreover depends logarithmically on the lattice
spacing. To see this explicitly, let us calculate the AC conductivity{\Large %
\ }approximately by using the Weyl approximation (linearization of the
dispersion relation) for the contribution of the Weyl points.{\Large \ }%
Within the continuum approximation in spherical coordinates one writes the
sum in Eq.(\ref{sigma_eta}) as:

\begin{eqnarray}
\sigma _{0}\left( \Omega \right) &=&-\frac{8ie^{2}}{\left( 2\pi \right) ^{3}}%
\int_{k=0}^{\Lambda }k\int_{\theta ,\phi }\sin ^{3}\theta \frac{\Omega }{%
4k^{2}-\Omega _{+}^{2}}  \label{sigma0a} \\
&=&-\frac{32e^{2}\Omega }{3\left( 2\pi \right) ^{2}}\int_{k=0}^{\Lambda }%
\frac{k}{4k^{2}-\Omega _{+}^{2}}\text{.}  \notag
\end{eqnarray}%
where the factor $8$ is due to the multiplicity of the Weyl points in our
"universal" tight binding model. The integration over $k$ finally gives:%
\begin{eqnarray}
\sigma _{0}\left( \Omega \right) &=&\frac{e^{2}\Omega }{3\pi }\left( 1-\frac{%
i}{\pi }\log \frac{4\Lambda ^{2}-\Omega ^{2}}{\Omega ^{2}}\right)
\label{sigma0final} \\
&\approx &\frac{e^{2}\Omega }{3\pi }-i\frac{e^{2}\Omega }{3\pi ^{2}}\log 
\frac{4\Lambda ^{2}}{\Omega ^{2}}\text{.}  \notag
\end{eqnarray}%
In the last line small terms of the relative order $\Omega ^{2}/\Lambda ^{2}$
were omitted. The finite result for the real part and regularized Matsubara
conductivity were first obtained in\cite{Vishwanath12}.

Let us emphasize that the electric properties in 3D differ from that in 2D
(graphene) in that there appears an imaginary part of conductivity that is
of the same size and sometimes even larger than the pseudo-dissipative one.
Now we turn to the study of the many-body effects in 3D Weyl semimetal.

\section{The interaction correction to the AC conductivity of the Weyl
semimetal}

\subsection{Leading corrections to the current evolution in a time dependent
field}

The expressions for the diamagnetic and the paramagnetic components of the
current density as a linear response to arbitrarily time dependent electric
fields are given by Eqs.(\ref{JpJd_def}) and (\ref{Jplinresp}) respectively.
The evolution operator however should be modified as 
\begin{equation}
e^{-i\left( K+V\right) t}=e^{-iKt}\left[ 1-i\int_{\tau =0}^{t}e^{iK\tau
}Ve^{-iK\tau }+O\left( V^{2}\right) \right] ,  \label{Evolution}
\end{equation}%
and the ground state (at initial time or Schr\"{o}dinger) as%
\begin{equation}
\left\vert \psi \left( 0\right) \right\rangle =\left\vert 0\right\rangle
+\left\vert \psi _{1}\right\rangle ;\text{ }\left\vert \psi
_{1}\right\rangle =-\frac{1}{K}V^{\left( 4,0\right) }\left\vert
0\right\rangle  \label{vaccumcor}
\end{equation}%
with the correction to energy being $\Delta E=\left\langle 0\left\vert
V\right\vert 0\right\rangle =0$, since the constant term in the energy will
be consistently omitted after the Hamiltonian was normal ordered. Only
creation of four particles at once is possible within the "universal" model.
as is shown in Appendix B. The pair creating part $V^{\left( 2,0\right) }$
vanishes. This simplifies the problem significantly. For example, the
diamagnetic component is absent due to this. Indeed the diamagnetic current
is quadratic, Eq(\ref{JpJd_def}), and hence the correction%
\begin{equation}
\Delta j^{d}\left( t\right) =\left\langle \psi _{1}\left\vert
e^{-iKt}j^{d}e^{iKt}\right\vert 0\right\rangle +c.c.=\left\langle \psi
_{1}\left\vert j^{d}\right\vert 0\right\rangle +c.c.=0\text{.}
\label{nodiag}
\end{equation}

The paramagnetic correction (in homogeneous electric field along certain
direction described by vector potential $A\left( t\right) $) of Eq.(\ref%
{Jplinresp}) in the presence of Coulomb interactions takes a form:

\begin{eqnarray}
j^{p}\left( t\right) &=&-ie^{2}\int_{t_{1}=0}^{t}A\left( t_{1}\right) \Xi
\left( t-t_{1}\right) +cc;  \label{deljp} \\
\Xi \left( t-t_{1}\right) &=&\left\langle \psi \left( 0\right) \left\vert
e^{iHt}j^{p}e^{-iH\left( t-t_{1}\right) }j^{p}e^{-iHt_{1}}\right\vert \psi
\left( 0\right) \right\rangle  \notag \\
&=&\left\langle \psi \left( 0\right) \left\vert j^{p}e^{-iH\left(
t-t_{1}\right) }j^{p}\right\vert \psi \left( 0\right) \right\rangle \text{.}
\notag
\end{eqnarray}%
Expanding in interaction with help of Eqs.(\ref{Evolution}) and (\ref%
{vaccumcor}), the correction to the expectation value in the interacting
ground state $\Xi $ above in terms of the unperturbed vacuum expectation
values is:%
\begin{widetext}%

\begin{equation*}
\Delta \Xi \left( t-t_{1}\right) =\ \ \ \ \ \ \ \ \ \ \ \ \ \ \ \ \ \ \ \ \
\ \ \ \ \ \ \ \ \ \ \ \ \ \ \ \ \ \ \ \ \ \ \ \ \ \ \ \ \ \ \ \ \ \ \ \ \ \
\ \ \ \ \ \ \ \ \ \ \ \ \ \ \ \ \ \ \ \ \ \ \ \ \ \ \ \ \ \ \ \ \ \ \ \ \ \
\ \ \ \ \ \ \ \ \ \ \ \ \ \ \ \ \ \ \ \ \ \ \ \ \ \ \ \ \ \ \ \ \ \ \ \ \ \
\ \ \ \ \ \ \ \ \ \ \ \ \ \ \ 
\end{equation*}

\begin{eqnarray}
&&-i\left\langle 0\left\vert j^{p}\int_{\tau =0}^{t-t_{1}}e^{-iK\left(
t-t_{1}-\tau \right) }\left( V^{\left( 1,1\right) }+V^{\left( 2,2\right)
}\right) e^{-iK\tau }j^{p}\right\vert 0\right\rangle -\left\langle
0\left\vert j^{p}e^{-iK\left( t-t_{1}\right) }j^{p}\frac{1}{K}V^{\left(
4,0\right) }\right\vert 0\right\rangle -\left\langle 0\left\vert V^{\left(
0,4\right) }\frac{1}{K}j^{p}e^{-iK\left( t-t_{1}\right) }j^{p}\right\vert
0\right\rangle \text{.}  \notag \\
&&  \label{ksidef}
\end{eqnarray}%
\end{widetext}%
The other contributions vanish, again due to absence of pair creations at
this order. The first term corresponds to the fermion self energy
correction, while the rest describe the vertex correction. These are readily
calculated:%
\begin{widetext}%

\begin{equation}
\Delta j\left( t\right) =\frac{e^{2}}{\mathcal{V}^{2}}\int_{t_{1}=0}^{t}A%
\left( t_{1}\right) \sum\limits_{\mathbf{pq}}v_{\mathbf{p-q}}\cdot \text{ \
\ \ \ \ \ \ \ \ \ \ \ \ \ \ \ \ \ \ \ \ \ \ \ \ \ \ \ \ \ \ \ \ \ \ \ \ \ \
\ \ \ \ \ \ \ \ \ \ \ \ \ \ \ \ \ \ \ \ \ \ \ \ \ \ \ \ \ \ \ \ \ \ \ \ \ \
\ \ \ \ \ \ \ \ \ \ \ \ \ \ \ \ \ \ \ \ \ \ }  \label{ksinew}
\end{equation}

\begin{equation}
\left\{ \left( t-t_{1}\right) \cos \left[ 2\varepsilon _{\mathbf{q}}\left(
t-t_{1}\right) \right] \left\vert \chi _{\mathbf{q}}\right\vert ^{2}g_{%
\mathbf{qp}}+g_{\mathbf{qp}}^{-}\frac{\sin \left[ 2\varepsilon _{\mathbf{q}%
}\left( t-t_{1}\right) \right] }{\varepsilon _{\mathbf{q}}+\varepsilon _{%
\mathbf{p}}}+\frac{1}{2}g_{\mathbf{qp}}^{+}\frac{\sin \left[ 2\varepsilon _{%
\mathbf{p}}\left( t-t_{1}\right) \right] -\sin \left[ 2\varepsilon _{\mathbf{%
q}}\left( t-t_{1}\right) \right] }{\varepsilon _{\mathbf{q}}-\varepsilon _{%
\mathbf{p}}}\right\} \text{,}  \notag
\end{equation}

\end{widetext}%
where $g_{\mathbf{qp}}$ is defined in Eq.(\ref{Aqp}) and 
\begin{equation}
g_{\mathbf{qp}}^{-}=Re\left[ \chi _{\mathbf{q}}^{\ast }\chi _{\mathbf{p}%
}^{\ast }\left( u_{\mathbf{p}}^{\ast }\cdot v_{\mathbf{q}}\right) ^{2}\right]
;\text{ \ }g_{\mathbf{qp}}^{+}=Re\left[ \chi _{\mathbf{q}}\chi _{\mathbf{p}%
}^{\ast }\left( v_{\mathbf{q}}^{\ast }\cdot v_{\mathbf{p}}\right) ^{2}\right]
\text{.}
\end{equation}%
This can be calculated using the same methods as in the leading order in the
previous section. The DC conductivity correction vanishes linearly. Finally
we calculate the correction to current for the AC electric field.

\subsection{Correction to the AC conductivity}

In the homogeneous AC electric field, $A\left( t\right) =-\frac{cE}{\Omega }%
\sin \left( \Omega t\right) $, the AC conductivity averaged over time as in
the leading order, Eq.(\ref{etatrick}), is:%
\begin{widetext}%

\begin{equation}
\sigma _{1}\left( \Omega \right) =\frac{2ie^{2}}{\mathcal{V}^{2}\Omega }%
\sum\limits_{\mathbf{pq}}\frac{v_{\mathbf{p-q}}}{4\varepsilon _{\mathbf{q}%
}^{2}-\Omega _{+}^{2}}\left[ -\frac{\left\vert \chi _{\mathbf{q}}\right\vert
^{2}g_{\mathbf{qp}}\left( 4\varepsilon _{\mathbf{q}}^{2}+\Omega ^{2}\right) 
}{4\varepsilon _{\mathbf{q}}^{2}-\Omega _{+}^{2}}+\frac{2g_{\mathbf{qp}%
}^{-}\varepsilon _{\mathbf{q}}}{\varepsilon _{\mathbf{q}}+\varepsilon _{%
\mathbf{p}}}+\frac{g_{\mathbf{qp}}^{+}\left( 4\varepsilon _{\mathbf{q}%
}\varepsilon _{\mathbf{p}}+\Omega ^{2}\right) }{4\varepsilon _{\mathbf{p}%
}^{2}-\Omega _{+}^{2}}\right] \text{,}  \label{sigmacorr1}
\end{equation}%
\end{widetext}%
where $\Omega _{+}\equiv \Omega +i\eta $. Subtracting the DC limit (that
vanishes after averaging over time) like in the leading order, one obtains a
much more converging expression proportional to the frequency:%
\begin{widetext}%

\begin{equation}
\sigma _{1}\left( \Omega \right) =\frac{ie^{2}\Omega }{2\mathcal{V}^{2}}%
\sum\limits_{\mathbf{pq}}\frac{v_{\mathbf{p-q}}}{4\varepsilon _{\mathbf{q}%
}^{2}-\Omega _{+}^{2}}\left[ -\frac{\left\vert \chi _{\mathbf{q}}\right\vert
^{2}g_{\mathbf{qp}}\left( 12\varepsilon _{\mathbf{q}}^{2}-\Omega ^{2}\right) 
}{\varepsilon _{\mathbf{q}}^{2}\left( 4\varepsilon _{\mathbf{q}}^{2}-\Omega
_{+}^{2}\right) }+\frac{2g_{\mathbf{qp}}^{-}}{\varepsilon _{\mathbf{q}%
}\left( \varepsilon _{\mathbf{q}}+\varepsilon _{\mathbf{p}}\right) }+\frac{%
g_{\mathbf{qp}}^{+}\left( 4\varepsilon _{\mathbf{q}}\varepsilon _{\mathbf{p}%
}+4\varepsilon _{\mathbf{q}}^{2}+4\varepsilon _{\mathbf{p}}^{2}-\Omega
^{2}\right) }{\varepsilon _{\mathbf{q}}\varepsilon _{\mathbf{p}}\left(
4\varepsilon _{\mathbf{p}}^{2}-\Omega _{+}^{2}\right) }\right] \text{.}
\label{sigcorr2new}
\end{equation}%
\end{widetext}%
The numerical evaluation results in (without subleading terms in the
imaginary part)%
\begin{equation}
\sigma _{1}\left( \Omega \right) =\frac{e^{4}\Omega }{9\pi ^{2}}\left[ -\log 
\frac{\pi ^{2}}{a^{2}\Omega ^{2}}-5+\frac{i}{8\pi }\log ^{2}\frac{\pi ^{2}}{%
a^{2}\Omega ^{2}}\right] \text{.}  \label{delsigexact}
\end{equation}%
It is however instructive to obtain this expression within the continuum
(Weyl) approximation valid within radius $\Lambda $ around each of the eight
Weyl points.

Near a Weyl point one has the following expansion of the functions appearing
in Eq.(\ref{sigcorr2new}) in spherical coordinates around one of the points
(chosen to be the origin $\Gamma $):%
\begin{eqnarray}
\chi _{\mathbf{k}} &=&-ie^{i\phi _{\mathbf{k}}}\sin \theta _{\mathbf{k}}
\label{hWeyl} \\
g_{\mathbf{qp}}^{\pm } &=&\frac{1}{2}\sin \theta _{\mathbf{q}}\sin \theta _{%
\mathbf{p}} \\
&&\cdot \left[ \cos \left( \phi _{\mathbf{q}}-\phi _{\mathbf{p}}\right)
\left( \cos \theta _{\mathbf{q}}\cos \theta _{\mathbf{p}}\pm 1\right) +\sin
\theta _{\mathbf{q}}\sin \theta _{\mathbf{p}}\right] .  \notag
\end{eqnarray}%
complementing Eq.(\ref{AqpWeyl}) for $g_{\mathbf{qp}}$. The sum and the
Coulomb potential in Eq.(\ref{sigcorr2new}) are written as%
\begin{widetext}
\begin{equation}
\frac{8}{2\left( 2\pi \right) ^{6}}\int_{pq\phi _{\mathbf{p}}\phi _{\mathbf{q%
}}\theta _{\mathbf{p}}\theta _{\mathbf{q}}}\frac{4\pi p^{2}q^{2}\sin \theta
_{\mathbf{p}}\sin \theta _{\mathbf{q}}}{p^{2}+q^{2}-2pq\left[ \cos \theta _{%
\mathbf{p}}\cos \theta _{\mathbf{q}}+\sin \theta _{\mathbf{p}}\sin \theta _{%
\mathbf{q}}\cos \left( \phi _{\mathbf{p}}-\phi _{\mathbf{q}}\right) \right] }%
.  \label{angleint}
\end{equation}%
\end{widetext}%
The integrals over all the angles can be performed, see Appendix B,%
\begin{widetext}%
\begin{eqnarray}
\sigma _{1}\left( \Omega \right) &=&\frac{2ie^{4}\Omega }{\left( 2\pi
\right) ^{3}}\int_{q,p=0}^{\Lambda }\frac{1}{4q^{2}-\Omega _{+}^{2}}\left\{ 
\frac{2G^{-}\left( r\right) }{1+r}+\frac{4pq+4q^{2}+4p^{2}-\Omega ^{2}}{%
\left( 4p^{2}+\Omega _{+}^{2}\right) }G^{+}\left( r\right) -\frac{%
2G^{+}\left( r\right) }{r}\frac{12q^{2}-\Omega ^{2}}{4q^{2}+\Omega _{+}^{2}}%
\right\}  \label{afterangle} \\
G &=&\frac{2\left( 1+r^{2}\right) }{3r}\log \frac{\left( 1+r\right) ^{2}}{%
\left( 1-r\right) ^{2}}-\frac{8}{3};\text{ \ }G^{\pm }=\frac{\left( 1\mp
r\right) ^{4}}{6r^{2}}\log \frac{\left( 1+r\right) ^{2}}{\left( 1-r\right)
^{2}}\pm 4-\frac{2\left( 1\pm r\right) ^{2}}{3r}\text{,}  \notag
\end{eqnarray}%
\end{widetext}%
with $r=q/p$. The integrals give

\begin{equation}
\sigma _{1}\left( \Omega \right) =\frac{e^{4}\Omega }{9\pi ^{2}}\left[ -\log 
\frac{4\Lambda ^{2}}{\Omega ^{2}}-5\,+\frac{i}{8\pi }\log ^{2}\frac{4\Lambda
^{2}}{\Omega ^{2}}\right] \text{.}  \label{Weyl res}
\end{equation}%
One observes that the subleading terms cannot be given correctly by the Weyl
approximation as expected. We can use the renormalized velocity from Eq.(\ref%
{RenV}) to make the real part finite..

\subsection{Renormalization of the perturbative expansion for conductivity}

We have calculated the renormalization of the Fermi velocity, Eq.(\ref{RenV}%
) and the AC conductivity of the clean Weyl semimetal within the leading
order in Coulomb interaction. It is very tempting to try to use
renormalization to improve the results of the "bare" perturbation theory
presented to the "two-loop" order above by renormalizing the parameters of
the theory. The expression for the AC conductivity both in the leading order
and for the interaction correction contains dependence on the UV cutoff $a$
or $\Lambda $ and it is interesting to ask whether physically measurable
quantities can be rewritten via "renormalized" parameters only or the
microscopic details represented by the cutoff dependence are indeed
unavoidable for certain measurable quantities like the AC conductivity.

The possibility of "renormalizability" is expected for the 3D Weyl model on
the following two grounds.

1. It has been claimed recently and shown to a very high order explicitly
that the 2D version (graphene) is renormalizable\cite{Gonzalez}. This is
surprising due to the breaking of the relativistic invariance (that ensures
the renormalizability for a 2D model with a 3D electromagnetic coupling).
The AC conductivity expression in 2D can be written via the renormalized
Fermi velocity, Eq.(\ref{RenV}) replacing the bare one in Eq.(\ref{Weyl res}%
). The key point here is that the leading order conductivity is independent
of both the velocity and the UV cutoff.

2. The relativistic version, QED, is renormalizable. Breaking of the
relativistic invariance by taking just the static part of the interaction
might not spoil this like in 2D.

We perform the renormalization of parameters up to the two-loop order.

Replacing the bare Fermi velocity with the renormalized one from Eq.(\ref%
{RenV}) in the real part of the conductivity from Eq.(\ref{results}), the
two-loop result becomes finite and proportional to frequency:%
\begin{widetext}%

\begin{eqnarray}
\sigma ^{\prime } &=&\frac{e^{2}\Omega }{3\pi v_{r}}\left( 1+\frac{\alpha }{%
3\pi }\log \frac{4v_{r}^{2}\Lambda ^{2}}{\overline{\Omega }^{2}}\right) +%
\frac{e^{2}\alpha \Omega }{9\pi ^{2}v_{r}}\left( -\log \frac{%
4v_{r}^{2}\Lambda ^{2}}{\Omega ^{2}}-5\,\right) \\
&=&\frac{e^{2}\Omega }{3\pi v_{r}}\left[ 1+\,\frac{\alpha }{3\pi }\left(
\log \frac{\Omega ^{2}}{\overline{\Omega }^{2}}+C\right) +O\left( \alpha
^{2}\right) \right] \text{.}  \notag
\end{eqnarray}%
\end{widetext}%
where $\alpha =\frac{e^{2}}{v_{r}\hbar }$ leading to the result (in physical
units) given in Eqs.(\ref{results}). As in 2D the constant $C$ is positive
and of order $1$. Note that the coefficient $e^{2}$ in the definition of the
current is not directly related by relativistic invariance to the Coulomb
interaction in the present model, and thus is not renormalized. In the order
considered there is no need to renormalize the static Coulomb coupling. The
imaginary part remains therefore "divergent", namely logarithmically
dependent on microscopic details.

\section{Discussion, experimental consequences}

Before discussing the applicability of the results to materials proposed as
realizations of the Weyl semimetal, let us summarize the electromagnetic
properties of the clean 3D semimetals at zero temperature that can be
extracted from the AC conductivity, Eqs.(\ref{sig0},\ref{sig0Im},\ref%
{results},\ref{Imsig}). Extension of our formulas to finite temperature is
trivial via the Matsubara substitution $\Omega \rightarrow \Omega -i\hbar
/k_{B}T$. The properties of the clean semimetal are expected to be
dissimilar from those of a band insulator and a metal. In particular, the
dielectric and optical properties differ markedly from both of them. The
complex conductivity of the neutral Weyl plasma at zero temperature is very
different from semiconductors and from metals. Let us contrast it with the
Lorentz model of a band insulator and the Drude model for metals with
electron density $n$ and relaxation time $\tau $. The AC conductivity of the
Lorentz model of a band insulator (semiconductor), represented in Fig. 3 by
the red line, is

\begin{equation}
\sigma _{ins}\left( \Omega \right) =\frac{\omega _{p}^{2}\tau }{4\pi }\frac{%
\Omega }{\Omega +i\tau \left( \omega _{0}^{2}-\Omega ^{2}\right) }\text{,}
\end{equation}%
where the central frequency of the band was taken to be rather small $\omega
_{0}=3\cdot 10^{14}Hz,$ the relaxation time $\tau =2\cdot 10^{-14}\sec $ and
the "plasma" frequency $\omega _{p}=10^{14}Hz$. The Drude conductivity of a
metal, represented by the blue line, is obtained from this formula by taking 
$\omega _{0}=0$ and the values of $\tau =2\cdot 10^{-14}\sec $ and $\omega
_{p}=2\cdot 10^{14}Hz$. These are compared with the real part and the
imaginary part, Figs.3a and 3b respectively, of the Weyl semimetal,
represented by the magenta line. The number of Weyl points is $N_{W}=8$, the
UV cutoff $\Lambda =\pi /a$, $a=3A$, $v=10^{6}m/s$ and the intrinsic
dielectric constant $\epsilon =3$ (due degrees of freedom not included in
the model). At low frequencies the absorptive part is linear, so that at DC
the Weyl semimetal is insulating. However it becomes comparable with that of
(a poor) metal at THz frequencies. The imaginary part is linear as in a
metal but with the opposite sign (capacitive like in insulator rather than
inductive).

\begin{figure}[tbp]
\begin{center}
\includegraphics[width=6cm]{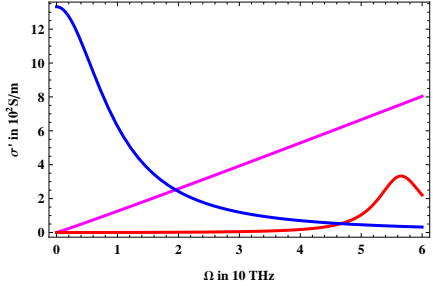}
\end{center}
\caption{Fig.3a The real part (Fig.3a)
and the imaginary part (Fig.3b) of the AC conductivity of the Weyl
semimetal, (magenta line) are compared with those of a band insulator (red
line) and of a metal (blue line). The parameter values are given in the text.}
\end{figure}

\begin{figure}[tbp]
\begin{center}
\includegraphics[width=6cm]{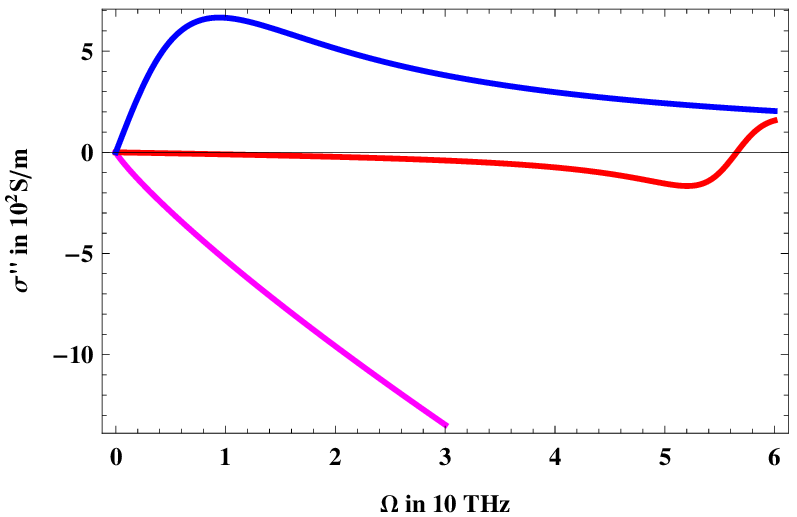}
\end{center}
\caption{Fig.3b} 
\end{figure}

The complex dielectric constant of the 3D semi-metal is $\varepsilon =1+4\pi
i\sigma /\Omega =\varepsilon ^{\prime }+i\varepsilon ^{\prime \prime }$ 
\begin{eqnarray}
\varepsilon ^{\prime }\left( \Omega \right) &=&1+\frac{N_{W}e^{2}}{3\pi
v\hbar }\log \frac{\Lambda v}{\Omega }\left[ 1+\frac{\alpha }{3\pi }\log 
\frac{\Lambda v}{\Omega }\right] ;  \label{eps} \\
\varepsilon ^{\prime \prime }\left( \Omega \right) &=&\frac{N_{W}e^{2}}{%
6v\hbar }\left[ 1+\,\alpha \left( \frac{2}{3\pi }\log \frac{\Omega }{%
\overline{\Omega }}+C\right) \right] \text{.}  \notag
\end{eqnarray}%
where $\alpha =e^{2}/\epsilon v\hbar $ with the renormalized (measured)
value of the Fermi velocity $v$. The normalization frequency is $\overline{%
\Omega }$ and $C=-\frac{5}{3\pi }\approx -0.53$. The real part of the
dielectric constant $\varepsilon ^{\prime }>1$ is like that of an ordinary 
\textit{dielectric} material with a very weak frequency dependence, see Fig.
4a, despite the nonzero AC conductivity which is linear in frequency. Albeit
note the logarithmic divergence for small frequencies: for example at $%
\Omega =100MHz$, $\varepsilon ^{\prime }=16$, see Fig.4a. The imaginary part
of the dielectric constant depends also only weakly on frequency and is
universal in that it is of order $0.2N_{W}$ for Fermi velocity $v\sim
10^{6}m/s$ and intrinsic dielectric constant $\epsilon =3$.

\begin{figure}[tbp]
\begin{center}
\includegraphics[width=6cm]{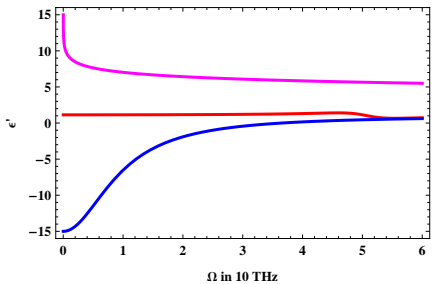}
\end{center}
\caption{Fig.4a The real part (Fig.4a)
and the imaginary part (Fig.4b) of the dielectric constant of the Weyl
semimetal, (magenta line) are compared with those of a band insulator (red
line) and of a metal (blue line). The parameter values are given in the text.}
\end{figure}

\begin{figure}[tbp]
\begin{center}
\includegraphics[width=6cm]{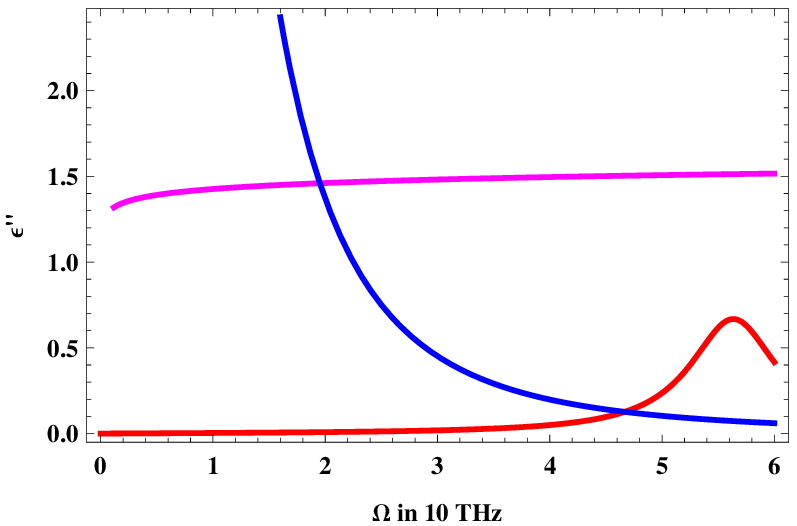}
\end{center}
\caption{Fig.4b} 
\end{figure}

Let us compare the electric and optical properties of the pure 3D Weyl
semimetal at neutrality point with the corresponding ones in 2D. In the pure
2D Weyl semimetal at the neutrality point (suspended undoped graphene is
considered to be a good realization of this model\cite{Andrei}) the real
part of the conductivity is finite and frequency independent in infinite
samples and with no contact work function\cite{Lewkowicz11}, and there is no
imaginary part in the AC conductivity, at least to leading order in
interactions. The transport therefore is purely pseudo-dissipative. As noted
before, the situation changes rather dramatically in 3D. We expect that
problems of separation between the interband transitions (between the
valence and the conduction bands or electron-hole pairs effects) considered
in the present study and the intraband transitions (including the Klein
scattering) due to potential barriers and mesoscopic effects are less
pronounced in 3D compared to 2D. Let us now discuss the limitations of the
model and point out some immediate extensions.

A clean system was assumed, while disorder is expected to be present. The
effects of disorder, neglecting the interactions, were studied in ref.\cite%
{Vishwanath12}. There might be an interplay between the disorder and
interaction effects, but the basic physics is expected to be unaltered
unless interaction or disorder are strong. One also can hope that, like in
graphene, the importance of disorder might be reduced compared to
expectations based on physics of ordinary "nonrelativistic" quasiparticles.
The use of an approximation like the tight binding model of section II or
even a continuum low energy Weyl model is justified as long as the
frequencies considered are much lower than the band width. As we argued in
section V, the tight binding model is rather universal, but the use of \ the
effective massless Weyl theory on the condensed matter scale has to be dealt
with care.

It is well known in field theory\cite{Smit} and in graphene that massless
fermions cause\ the absence of a perfect scale separation between high
energies (on atomic scale $\gamma $) and low energies (effective Weyl theory
on the condensed matter scale $<<\gamma $). It was demonstrated in the
context of graphene\cite{Kao} that some aspects of the linear response
physics, including the Coulomb interactions corrections\cite{Rosenstein13},
are \textit{not} dominated by the Weyl points of the Brillouin zone at which
the effective low energy model is valid. For example, large contributions
(infinite, when the size of the Brillouin zone is being considered infinite)
to the conductivity from the vicinity of the Weyl points are cancelled by
contributions from the region between them. Another famous consequence of
this scale nonseparation is the "species doubling" of lattice fermions\cite%
{Smit}, which in the context of graphene means that there necessarily
appears a pair of Weyl points of opposite chirality. The UV regularization
of the effective theory \textit{does matter} and, if one were to use such a
model, the only regularization known to date to be consistent with the tight
binding is the space dimensional regularization developed in ref. \cite%
{Herbut2}.

There are a number of quite straightforward extensions of the leading order
interaction calculation of the AC conductivity performed in the present
work. Converting it into the RPA - like approximation is simple, but in
addition taking into account the finite momentum transfer is more involved
than the analogous calculations in graphene\cite{MacDonald}. This would
allow to study the plasmons' effects and even the strong coupling phenomena
like the exciton condensation\cite{exciton3D}. Beyond the linear response,
phenomena like the nonlinear I-V curves due to the particle-hole (Schwinger)
pair creation and relaxation due to their recombination studied in the
context of graphene\cite{nonlinear} also can be extended to 3D in a
straightforward fashion.

Acknowledgements. We are indebted to T. Maniv, H.C. Kao and W.B. Jian for
valuable discussions.

\section{Appendix A. The universal tight binding model}

\subsection{Diagonalization of the noninteracting tight binding model and
expression for the current}

The noninteracting model is diagonalized in 3D with the following
coefficients of the Bogoliubov transformation Eq.(\ref{Bogtrans}):

\begin{eqnarray}
u_{\mathbf{k}} &=&\frac{1}{\sqrt{2\varepsilon _{k}\left( \varepsilon _{k}+%
\widehat{k}_{z}\right) }}%
\begin{pmatrix}
-\widehat{k}_{x}+i\widehat{k}_{y} \\ 
\varepsilon _{k}+\widehat{k}_{z}%
\end{pmatrix}%
;\ \   \label{uv} \\
v_{\mathbf{k}} &=&\frac{1}{\sqrt{2\varepsilon _{k}\left( \varepsilon _{k}+%
\widehat{k}_{z}\right) }}%
\begin{pmatrix}
\varepsilon _{k}+\widehat{k}_{z} \\ 
\widehat{k}_{x}+i\widehat{k}_{y}%
\end{pmatrix}%
,  \notag
\end{eqnarray}%
where $\widehat{k}\equiv \sin k$ and $\varepsilon _{k}^{2}=\widehat{k}%
_{x}^{2}+\widehat{k}_{y}^{2}+\widehat{k}_{z}^{2}$. Using matrix elements

\begin{equation}
u_{k}^{+}\sigma _{3}v_{k}=-\frac{\widehat{k}_{x}+i\widehat{k}_{y}}{%
\varepsilon _{k}};\text{ \ \ \ }v_{k}^{+}\sigma _{3}v_{k}=-u_{k}^{+}\sigma
_{3}u_{k}=\frac{\widehat{k}_{z}}{\varepsilon _{k}},  \label{matrixelements}
\end{equation}%
one obtains the coefficient of the electric current as%
\begin{equation}
\chi _{\mathbf{k}}=\frac{\left( \widehat{k}_{y}-i\widehat{k}_{x}\right) \cos
k_{z}}{\varepsilon _{k}};\text{ \ \ }\iota _{\mathbf{k}}=\frac{\widehat{k}%
_{z}\cos k_{z}}{2\varepsilon _{k}}\text{.}  \label{chidef}
\end{equation}

\subsection{Normal ordering of the interaction}

The normal ordering is quite straightforward.

\begin{equation}
V^{ij}=\sum\limits_{klk^{\prime }l^{\prime }}v_{klk^{\prime }l^{\prime
}}O^{ij}\text{,}  \label{general}
\end{equation}%
where the part $V^{ij}$ contains $i$ creation operators $a^{+}$ or $b^{+}$
and $j$ annihilation operators $a$ or $b$. The quartic terms are:%
\begin{widetext}%

\begin{equation}
O^{40}=-\left( v_{l}^{\ast }\cdot u_{k}\right) \left( v_{l^{\prime }}^{\ast
}\cdot u_{k^{\prime }}\right) a_{l}^{+}a_{l^{\prime
}}^{+}b_{-k}^{+}b_{-k^{\prime }}^{+}=\left( O^{04}\right) ^{+};  \label{O40}
\end{equation}

\begin{eqnarray}
O^{31} &=&\left( O^{13}\right) ^{+}=\left( v_{l}^{\ast }\cdot u_{k}\right) 
\left[ \left( v_{l^{\prime }}^{\ast }\cdot v_{k^{\prime }}\right)
a_{l}^{+}b_{-k}^{+}a_{l^{\prime }}^{+}a_{k^{\prime }}-\left( u_{l^{\prime
}}^{\ast }\cdot u_{k^{\prime }}\right) a_{l}^{+}b_{-k}^{+}b_{-k^{\prime
}}^{+}b_{-l^{\prime }}\right]  \label{O31} \\
&&+\left( v_{l^{\prime }}^{\ast }\cdot u_{k^{\prime }}\right) \left[ \left(
v_{l}^{\ast }\cdot v_{k}\right) a_{l}^{+}a_{l^{\prime }}^{+}b_{-k^{\prime
}}^{+}a_{k}-\left( u_{l}^{\ast }\cdot u_{k}\right) \right] a_{l^{\prime
}}^{+}b_{-k}^{+}b_{-l}b_{-k^{\prime }}^{+};  \notag
\end{eqnarray}%
\begin{eqnarray}
O^{22} &=&-\left( v_{l^{\prime }}^{\ast }\cdot v_{k^{\prime }}\right) \left(
v_{l}^{\ast }\cdot v_{k}\right) a_{l}^{+}a_{l^{\prime
}}^{+}a_{k}a_{k^{\prime }}+\left( u_{l}^{\ast }\cdot u_{k}\right) \left(
v_{l^{\prime }}^{\ast }\cdot v_{k^{\prime }}\right) a_{l^{\prime
}}^{+}b_{-k}^{+}a_{k^{\prime }}b_{-l}+\left( u_{l^{\prime }}^{\ast }\cdot
u_{k^{\prime }}\right) \left( v_{l}^{\ast }\cdot v_{k}\right)
a_{l}^{+}b_{-k^{\prime }}^{+}a_{k}b_{-l^{\prime }}  \label{O22} \\
&&-\left( u_{l}^{\ast }\cdot u_{k}\right) \left( u_{l^{\prime }}^{\ast
}\cdot u_{k^{\prime }}\right) b_{-k}^{+}b_{-k^{\prime
}}^{+}b_{-l}b_{l^{\prime }}-\left( v_{l}^{\ast }\cdot u_{k}\right) \left(
u_{l^{\prime }}^{\ast }\cdot v_{k^{\prime }}\right)
a_{l}^{+}b_{-k}^{+}a_{k^{\prime }}b_{-l^{\prime }}-\left( u_{l}^{\ast }\cdot
v_{k}\right) \left( v_{l^{\prime }}^{\ast }\cdot u_{k^{\prime }}\right)
a_{l^{\prime }}^{+}b_{-k^{\prime }}^{+}a_{k}b_{-l}\text{.}  \notag
\end{eqnarray}%
\end{widetext}%
In principle there are three possible quadratic terms where%
\begin{widetext}%

\begin{equation}
O^{20}=\left( O^{02}\right) ^{+}=\delta _{k-l^{\prime }}\left( v_{l}^{\ast
}\cdot v_{k}\right) \left( v_{l^{\prime }}^{\ast }\cdot u_{k^{\prime
}}\right) a_{l}^{+}b_{-k^{\prime }}^{+}-\delta _{l-k^{\prime }}\left(
u_{l}^{\ast }\cdot u_{k}\right) \left( v_{l^{\prime }}^{\ast }\cdot
u_{k^{\prime }}\right) a_{l^{\prime }}^{+}b_{-k}^{+};  \notag
\end{equation}

\begin{eqnarray}
O^{11} &=&\delta _{k-l^{\prime }}\left( v_{l^{\prime }}^{\ast }\cdot
v_{k^{\prime }}\right) \left( v_{l}^{\ast }\cdot v_{k}\right)
a_{l}^{+}a_{k^{\prime }}+\delta _{l-k^{\prime }}\left( u_{l}^{\ast }\cdot
u_{k}\right) \left( u_{l^{\prime }}^{\ast }\cdot u_{k^{\prime }}\right)
b_{-k}^{+}b_{-l^{\prime }}  \label{O11} \\
&&-\left( u_{l}^{\ast }\cdot v_{k}\right) \left( v_{l^{\prime }}^{\ast
}\cdot u_{k^{\prime }}\right) \left( \delta _{k-l^{\prime }}b_{-k^{\prime
}}^{+}b_{-l}+\delta _{l-k^{\prime }}a_{l^{\prime }}^{+}a_{k}\right) .  \notag
\end{eqnarray}%
\end{widetext}%
Relations between the scalar products for unequal momenta that appear here

\begin{equation}
\left( u_{k}^{\ast }\cdot u_{l}\right) =\left( v_{k}^{\ast }\cdot
v_{l}\right) ^{\ast };\text{ \ \ }\left( v_{k}^{\ast }\cdot u_{l}\right)
=-\left( u_{k}^{\ast }\cdot v_{l}\right) ^{\ast }=-\left( v_{l}^{\ast }\cdot
u_{k}\right)  \label{relations}
\end{equation}%
are useful in summation over momenta incorporating Eq.(\ref{vdef}), in
particular for two equal momenta 
\begin{equation}
v_{klk^{\prime }k}=\frac{1}{\mathcal{V}}v_{l-k}\delta _{l-k^{\prime }};\ \ \
v_{kk^{\prime }k^{\prime }l^{\prime }}=\frac{1}{\mathcal{V}}v_{k^{\prime
}-k}\delta _{l^{\prime }-k}.  \label{particular}
\end{equation}%
Most importantly, summing over momenta, the pair creation term vanishes:%
\begin{widetext}%

\begin{eqnarray}
V^{20} &=&\sum\limits_{kl}v_{l-k}\left( -\left( u_{l}^{\ast }\cdot
v_{k}\right) \left( v_{k}^{\ast }\cdot v_{l}\right)
b_{-l}^{+}a_{l}^{+}+\left( u_{l}^{\ast }\cdot v_{k}\right) \left(
u_{k}^{\ast }\cdot u_{l}\right) b_{-k}^{+}a_{k}^{+}\right)  \label{V20zero}
\\
&=&\sum\limits_{kl}v_{l-k}\left( u_{l}^{\ast }\cdot v_{k}\right) \left(
\left( u_{l}^{\ast }\cdot u_{k}\right) -\left( v_{k}^{\ast }\cdot
v_{l}\right) \right) b_{-l}^{+}a_{l}^{+}=0\text{.}  \notag
\end{eqnarray}%
\end{widetext}%
This leads to numerous simplifications.

\section{Appendix B. Angle integrals appearing in calculations of self
energy and conductivity}

\subsection{Self energy}

The integral over $\phi _{\mathbf{q}}$ in Eq.(\ref{depsWeyl}) for the self
energy is, using%
\begin{widetext}
\begin{equation}
\int_{\Delta =0}^{2\pi }\frac{c+d\cos \Delta }{a+b\cos \Delta }=\frac{2\pi }{%
b}\left[ \left( bc-ad\right) /\sqrt{a^{2}-b^{2}}+d\right] ,  \label{deltaint}
\end{equation}%
is:%
\begin{eqnarray}
&&\frac{1}{2\pi }\int_{\phi _{q}=0}^{2\pi }\frac{\cos \theta _{q}\cos \theta
_{p}+\sin \theta _{q}\sin \theta _{p}\cos \left( \phi _{p}-\phi _{q}\right) 
}{1+r^{2}-2r\left[ \cos \theta _{p}\cos \theta _{q}+\sin \theta _{p}\sin
\theta _{q}\cos \left( \phi _{p}-\phi _{q}\right) \right] }  \label{fi_int}
\\
&=&\frac{1}{2r}\left[ \frac{1+r^{2}}{\sqrt{\left( 1+r^{2}-2r\cos \theta
_{p}\cos \theta _{q}\right) ^{2}-\left( 2r\sin \theta _{p}\sin \theta
_{q}\right) ^{2}}}-1\right] .  \notag
\end{eqnarray}%
\end{widetext}%
The integral over the azimuth angle for $\theta _{p}=0$ (our choice for the
quasiparticle direction) results in 
\begin{eqnarray}
&&\frac{1}{2\pi }\int_{\theta _{q}=0}^{\pi }\sin \theta _{q}\left( \frac{%
1+r^{2}}{1+r^{2}-2r\cos \theta _{q}}-1\right)  \label{Qpthetaint} \\
&=&\frac{1+r^{2}}{2r^{2}}\log \frac{\left( 1+r\right) ^{2}}{\left(
1-r\right) ^{2}}-\frac{2}{r},  \notag
\end{eqnarray}%
leading to Eq.(\ref{deps1}).

\subsection{Corrections to conductivity}

First contribution to the AC conductivity, Eq.(\ref{sigcorr2new}), involves
the angle integral%
\begin{widetext}%

\begin{equation}
I_{1}=\int_{\phi _{p}\phi _{q}\theta _{p}\theta _{q}}\sin \theta _{p}\sin
\theta _{q}\frac{\sin ^{2}\theta _{q}\left[ \cos \theta _{q}\cos \theta
_{p}+\sin \theta _{q}\sin \theta _{p}\cos \left( \phi _{p}-\phi _{q}\right) %
\right] }{p^{2}+q^{2}-2pq\left[ \cos \theta _{p}\cos \theta _{q}+\sin \theta
_{p}\sin \theta _{q}\cos \left( \phi _{p}-\phi _{q}\right) \right] }.
\label{I1}
\end{equation}%
Integration over $\phi _{p}$ and $\Delta \phi =\phi _{p}-\phi _{q}$ using
Eq.(\ref{deltaint}) gives

\begin{eqnarray}
I_{1} &=&\frac{2\pi ^{2}}{r}\int_{\theta _{p}\theta _{q}}\sin \theta
_{p}\sin ^{3}\theta _{q}\left[ \frac{1+r^{2}}{\sqrt{\left( 1+r^{2}-2r\cos
\theta _{p}\cos \theta _{q}\right) ^{2}-\left( 2r\sin \theta _{p}\sin \theta
_{q}\right) ^{2}}}-1\right]  \label{I1final} \\
&=&\frac{2\pi ^{2}}{r}G^{+}\left( r\right) ,  \notag
\end{eqnarray}%
\end{widetext}%
where function $G^{+}$ is given in Eq.(\ref{afterangle}). The two other
angle integrals are done in the same way.

\end{document}